\documentclass[aip,rsi,superscriptaddress,twocolumn]{revtex4-1}
\usepackage{graphicx} 
\graphicspath{{graph/}}
\usepackage[usenames]{color}
\usepackage{amsmath,amssymb}
\usepackage{gensymb}
\usepackage{natbib}
\usepackage{xcolor}
\def\strutdepth{\dp\strutbox}
\def\nw#1{\strut\vadjust{\kern-\strutdepth\vtop to0pt{\vss\hbox to\hsize
{\hskip\hsize\hskip5pt$\leftarrow$\hss\strut}}}{\em #1}}

\usepackage{rotating}
\usepackage{multirow}

\begin{document}

\title{A flexible rheometer design to measure the visco-elastic response of soft solids over a wide range of frequency}
\author{Etienne Rolley}
\affiliation{Laboratoire de Physique Statistique, UMR 8550 ENS-CNRS, Univ. Paris-Diderot, 24 rue Lhomond, 75005, Paris.}
\author{Jacco H. Snoeijer}
\affiliation{Physics of Fluids Group, Faculty of Science and Technology, University of Twente, P.O. Box 217, 7500 AE Enschede, The Netherlands}
\author{Bruno Andreotti}
\affiliation{Laboratoire de Physique Statistique, UMR 8550 ENS-CNRS, Univ. Paris-Diderot, 24 rue Lhomond, 75005, Paris.}

\begin{abstract}
We present a flexible set-up for determining the rheology of visco-elastic materials which is based on the mechanical response of a magnet deposited at the surface of a slab of material and excited electromagnetically. An interferometric measurement of the magnet displacement allows one to reach an excellent accuracy over a wide range of frequency. Except for the magnet, there is no contact between the material under investigation and the apparatus. At low frequency, inertial effects are negligible so that the mechanical response, obtained through a lock-in amplifier, directly gives the material complex modulus. At high frequency, damped waves are emitted and the rheology must be extracted numerically from a theoretical model. To validate the design, the instrument was used to measure the rheology of a test PDMS gel which presents an almost perfect scale free response at high frequency.
\end{abstract}

\date{\today}

\maketitle

Measuring the visco-elastic response of soft solids (i.e. non polycristalline) is necessary to understand their structure \cite{chen2010rheology}. As such, rheology is used to characterize the mechanical behaviour of a large variety of systems: reticulated polymers, gels \cite{miri2011viscosity}, biological tissues \cite{mijailovic2018characterizing}, etc. The need to explore rheological properties in a large frequency range and down to micro- or nano-scale \cite{cicuta2007microrheology} has recently triggered many efforts to design new setups using for instance Surface Force Apparatus\cite{cottin2002nanorheology,garcia2016micro}, various piezoelastic oscillators \cite{crassous2005characterization, belmiloud2008}, or ultrasonic wave propagation \cite{Lefebvre2018}. 
However, in many applications, rheological properties are still measured with conventional shear rheometers in which a flat sample is held between two plates. Measuring both the mobile plate displacement and the applied  force yields the complex shear modulus $\mu = G'+i G''$, where $G'(\omega)$ is the storage modulus and $G''(\omega)$ the loss modulus \cite{banks2011brief}. Accurate measurement requires a perfect control and determination of the sample geometry which has to match the plates of the rheometer, as well as a perfect contact between the sample and the plates.

Here, we present the principle of a rheometer where the sample is submitted to indentation rather than shear. Indentation has been widely used to characterize static or transient mechanical properties of solids \cite{sneddon1946boussinesq,zheng2017indentation} but not to determine the frequency dependence of the rheological response. In our setup, a magnetic oscillatory force is imposed on the indenter which is a permanent magnet. The magnet displacement is measured optically: there is no contact between the apparatus and the sample. The magnet inertia is small so that its displacement can be measured over 7 decades in frequency, up to $10$~kHz. As shown below, an instrument can be buit easily with simple and mostly off-the-shelve components. We derive the theoretical equations of the dynamical system, necessary to obtain the intrinsic shear modulus $\mu$ from the raw mechanical response. As an example, we work out the detailed response of silicone gel samples in various geometries.
\begin{figure}
	\centering
		\includegraphics{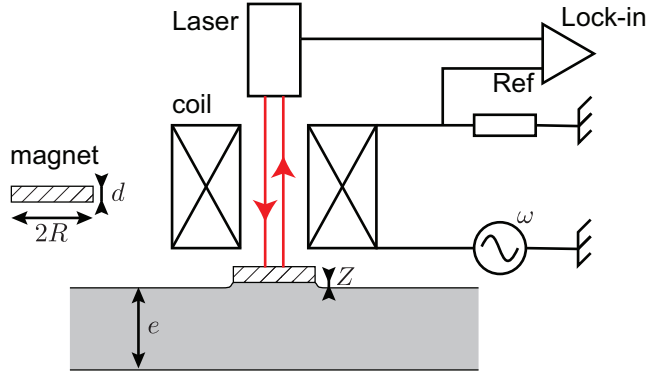}	
		\caption{Sketch of the rheometer: the magnetic force on the magnet is proportional to the current, the amplitude and phase of the displacement of the magnet is measured with a laser vibrometer and a lock-in amplifier.}
		\label{setup}
\end{figure}

\textit{Set-up~--~} The indenter is a small magnetic disc of radius $R$, which is placed on the surface of the soft solid under study. A magnetic force $F$ is applied with a coil positioned above the magnet (see fig. \ref{setup}). $F$ is proportional to the current $I$ in the coil, so that the force calibration can be easily performed with the magnet on a precision scale and operating the coil with DC current. In order to avoid electrical resonances and to ensure that $F$ and $I$ are proportional up to 10 kHz, it is important to minimize the stray capacitance between neighboring wires in the coil. To build the coil, we used a copper wire with a $0.2$ mm thick insulation, wrapped around a hollow cylinder. The vertical displacement $Z$ of the magnet is measured here with a commercial laser-interferometric vibrometer (SP-S model from SIOS Messtechnik GmbH), within a nominal resolution of $20\,\rm{pm}$.

One can infer from dimensional analysis that in the static limit, for a sample much larger than $R$ in all directions, the linear response should obey the scaling law: $F\sim \mu_0 R Z$, where $\mu_0=G'(\omega=0)$ is the static shear modulus. We therefore define the raw output of the experiment as the complex effective modulus $K(\omega)=K'(\omega)+iK''(\omega)$:
	\begin{equation}
		K \equiv \frac{1}{R} \frac{F}{Z}.
		\label{def}
	\end{equation}
$K(\omega)$ is measured using a digital lock-in amplifier using the intensity in the coil, measured with a shunt resistor, as a reference signal. Below we demonstrate how this measured quantity can be converted to the actual rheology $\mu(\omega)$ of the sample. The in-phase signal $K'(\omega)$ characterises the conservative part of the system (stiffness and inertia). The quadrature signal $K''(\omega)$ reflects dissipative processes.

This design offers a great flexibility, since the force range can be changed easily by modifying the coil, the magnitude of the current excitation, or the size of the magnet. In order to use sample sizes in the range 1-10 cm and to avoid strong finite size effects, Nd magnets with a radius $R$ from $1$ to $5$ mm, and a thickness around $0.5$ mm constitute a rational choice. This sets the typical size of the coil in the cm range. In our case, the inner and outer diameters of the coil  are respectively about $10$ and $25$ mm, and its height about $15$ mm. With $\sim 500$ windings the typical value for $F/I$ is then 0.01 N/A. In order to induce a measurable displacement with such a small force (under, say $I=1$ A, the typical elastic modulus must smaller than $1\, \rm{MPa}$, a range that covers many usual soft solids). Measurement reproducibility requires a good alignment of the coil and the magnet, and a fixed distance between these two elements. For our setup, a $1 \%$ reproducibility requires a positioning accuracy of about $0.1$ mm. 

As a test material, we have used a soft PDMS gel  (Dow Corning CY52-276, prepared in 1:1 ratio and cured at room temperature during 24h) for which the static modulus $\mu_0$ is of the order of $1$ kPa and which is nearly incompressible (the Poisson's ration $\nu \simeq 0.5$). For such a soft gel, a standard AC generator delivers a high enough current ($0.1$ A peak) to get a $100 {\rm \mu m}$ displacement at vanishing frequency. In the experiment reported here, the displacement $Z$ is measured with a laser vibrometer, offering a high sensitivity and a large bandwidth. Others schemes can be envisioned, including simple optical imaging, as the magnet can be visualized from the side. We emphasize that the use of a laser vibrometer makes the alignment of coil and magnet very easy. 

\textit{Overdamped limit~--~}We first focus on the overdamped case, where the inertia of the magnet and of the gel can be neglected. In this limit one can describe the sample's dynamical response to the disc-indentation using the formulas of static linear elasticity; the dynamical response is simply obtained by replacing the static shear modulus $\mu_0$ by the frequency-dependent $\mu(\omega)$. For thick samples $e \gg R$, we can thus consider the static solution of a disc that indents a semi-infinite medium, for which the normal stress $\sigma$ below the indenter reads~\cite{sneddon1946boussinesq}
\begin{equation}
\sigma=-\frac{2\mu}{\pi (1-\nu) \sqrt{R^2-r^2}} Z,
\end{equation}
were, $r$ is the radial distance from the origin. Neglecting solid capillary effects, the normal stress $\sigma$ vanishes outside the contact with the magnet ($r>R$). The force between the magnet and the sample is found by integration of the stress, and gives
\begin{equation}\label{eq:thick}
K = \frac{F}{RZ} =\frac{4 \mu}{1-\nu}.
\end{equation}
This provides the ``conversion factor" between the effective modulus $K$ (i.e. the scaled force measurement) and the intrinsic rheology $\mu$, for the case of thick samples. In many practical cases, however, samples are available only in the from of layers of thickness $e$ small compared to $R$. In this opposite limit, it is simpler to consider the incompressible case $\nu=1/2$, in order to apply the lubrication approximation for which the normal stress $\sigma$ at the free surface obeys the momentum balance (see Appendix):
\begin{equation}
\frac{\partial^2 \sigma}{\partial r^2}+\frac{1}{r}\;\frac{\partial \sigma}{\partial r}=\frac{3\mu}{e^3}Z.
\end{equation}
This can be integrated to 
\begin{equation}
\sigma=\frac{3\mu}{4e^3}(r^2-R^2) Z,
\end{equation}
which gives a force-displacement relation
\begin{equation}\label{eq:thin}
K=\frac{F}{RZ}=\frac{3\pi \mu R^3}{8e^3}.
\end{equation}
%
Equations~(\ref{eq:thick},\ref{eq:thin}) provide the asymptotic results, respectively, for large and small ratio $e/R$. For the general case of arbitrary thickness, we define the conversion function $\kappa(e/R)$ as
\begin{equation}
K = \mu \, \kappa(e/R).
\end{equation}
\begin{figure}
	\centering
		\includegraphics{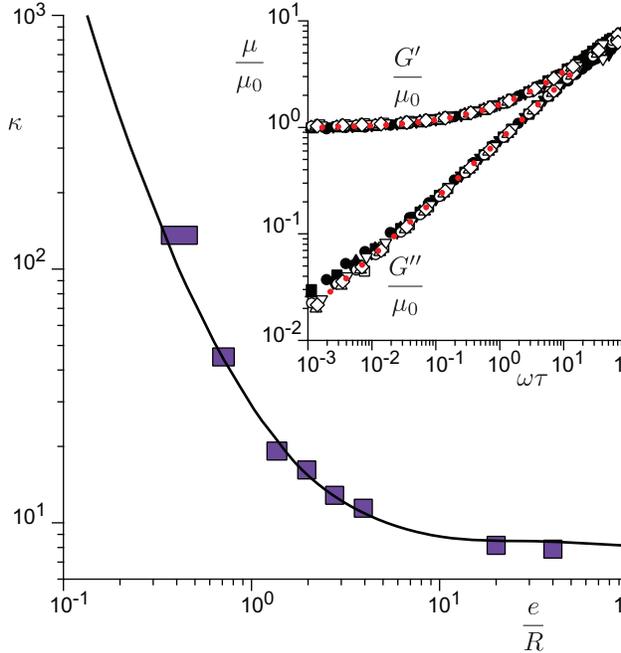}
		\caption{Scaling factor $\kappa=K/\mu$ as a function of the aspect ratio $e/R$. Dots: experimental values obtained from the rescaling of $K(\omega)$. The symbol size reflects the experimental error bars. Line: theoretical prediction in the overdamped limit (Poisson ratio $\nu=1/2$). Inset: rescaled rheology $\mu(\omega)$ for the different ratio $e/R$, superimposed to that obtained in a conventional Anton-Paar rheometer (red dots).}
		\label{finite}
\end{figure}

This formula has been  determined by numerical solution of the (incompressible) elastic problem for arbitrary thickness, and is shown as the solid line in fig.~\ref{finite}.
 In order to check our setup and to measure $\kappa$, we have performed systematic measurements of $K(\omega)$ at low frequency, for $R$ in the range $1$ to $5$ mm and $e$ in the range $1$ to $100$ mm. First, we verify that the rheological data $K(\omega)$ for various $e/R$ can be collapsed to a single curve (inset of fig. \ref{finite}). This confirms that, in the non-inertial regime, the same $\kappa$ applies to all frequencies. In addition, the scaled data  coincides with an independent measurement of $\mu(\omega)$ using  an Anton-Paar rheometer, up to the frequency $f=100$ Hz ($\omega\simeq 628$ rad/s). For the PDMS gel considered, the complex modulus is very accurately described\cite{ChambonWinter, LAL96,deGennes1996,ScanlanWinter} by $\mu(\omega)=\mu_0 \left[1+(i\omega\tau)^n\right]$, with $\mu_0=1.3$ kPa, $\tau = 0.13$ s and $n=0.55$. Our data are perfectly fitted with the same law for $\mu(\omega)$, with small variations of $\mu_0$ and $\tau$ depending on the curing procedure and the age of the sample. To complete the comparison, we first verified for a sample much larger than the magnet in all directions ($e/R \gg 1$) that $\kappa \simeq 8$, as expected from (\ref{eq:thick}) in the incompressible limit. Then, we have systematically analysed the results to samples of finite thickness. As shown in fig. \ref{finite}, the experimental data are in good agreement with the theoretical curve, determined numerically.

In the above calculation, we have neglected any contribution of capillary forces, although adhesion may be present in the experiment. The relative contribution of adhesion and elasticity is given by the dimensionless number $\mu_0 R/\gamma$, where $\gamma$ stands for the typical solid surface tension. This elasto-capillary number compares the typical elastic stress to the Laplace pressure: for a large enough magnet, i.e. for $R\gg \gamma/\mu_0$, adhesive effects can be safely neglected. Here, the elasto-cappilary length $\gamma/\mu_0$ is around  $10{\rm \mu m}$, which is 3 orders of magnitude smaller than the magnet radius.

\begin{figure}
	\centering
		\includegraphics{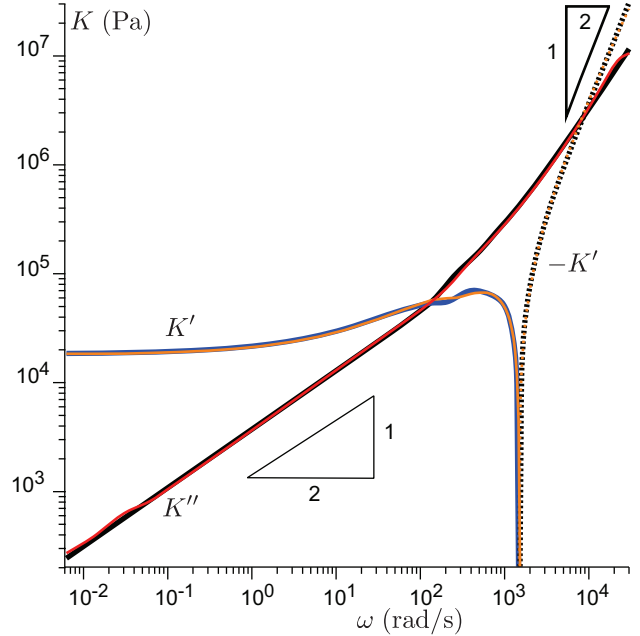}
		\caption{In-phase $K'$ and out-of-phase $K"$ response as a function of the angular frequency $\omega$ for the PDMS gel. By convention, the quantity is plotted with a solid line when positive and, when negative, its opposite is plotted in dotted line. Orange and red lines: measurements for a sample of PDMS gel much larger than the magnet ($R=5$ cm). Blue and black lines: prediction for $K'$ and $K"$ assuming that the gel shear modulus can be approximated by $\mu=\mu_0 \left[1+(i\omega\tau)^n\right]$.}
		\label{K}
\end{figure}

\textit{Inertial behavior~--~} We now turn to the dynamical behavior of the system at high frequency, for which inertial effects have to be taken into account. Here we consider only large samples to avoid any finite-size effect. Measurements of the effective modulus $K$ are shown in figure \ref{K} in the full angular frequency range, from $\omega=5\,\cdot 10^{-3}\,{\rm rad/s}$ to $2\,\cdot10^4\,{\rm rad/s}$.

An obvious feature of $K(\omega)$ is a resonance at an angular frequency $\omega_R$ around $1$ kHz, which is followed by a change of sign of the effective storage modulus $K'$. At angular frequencies well below the resonance, the effective modulus $K(\omega)$ is simply proportional to the elastic modulus $\mu(\omega)$, while above $\omega_R$ the oscillating magnet excites damped elastic waves inside the sample. Hence, the conversion from $K(\omega)$ to $\mu(\omega)$ becomes more intricate, as it involves the gel's inertia More precisely, at large frequency, an effective mass scaling as $\sim \rho_g R^3$  is set into an oscillatory motion, leading to an inertial term in $K'$ scaling as $- \rho_g R^2 \omega^2$. The asymptotic behaviour in $\omega^2$ is clearly visible in figure~\ref{K} (with a prefactor of approximately 1.6). The resonance frequency coincides with the cross-over from the quasi-static regime to the inertial regime and therefore is expected to scales as
\begin{equation}
\omega_R\sim \frac{1}{R}\;\sqrt{\frac{G'_R}{\rho_g}},
\end{equation}
where $G'_R$ is the storage modulus at the resonance frequency. This is indeed consistent with the experimental $\omega_R$, when using a multiplicative factor of about 1.5.

In addition, the inertia due to the magnet must be taken into account to extract the force. Namely, the force $F$ entering the definition of $K$ in (\ref{def}) is the force acting on the gel, which is obtained from the total force acting on the magnet by adding $-M_m \omega^2 Z$. This correction involves the magnet mass $M_m=\pi \rho_m R^2 d$, which is proportional to the density  $\rho_m \simeq 7.5 \;\mathrm{g/cm^3}$and to the thickness $d$. In order for the gel inertia to dominate over the magnet inertia, one needs $d/R $ to be much smaller than $\rho_g/\rho_m$. This condition is marginally realized in the example of figure (\ref{K}), for which we have choosen a small aspect ratio $d/R=0.08$ ($R=5$ mm and $d=0.4$ mm). In such a situation, the accuracy of the measurement is not affected when the correction is substracted.
\begin{figure}
	\centering
		\includegraphics{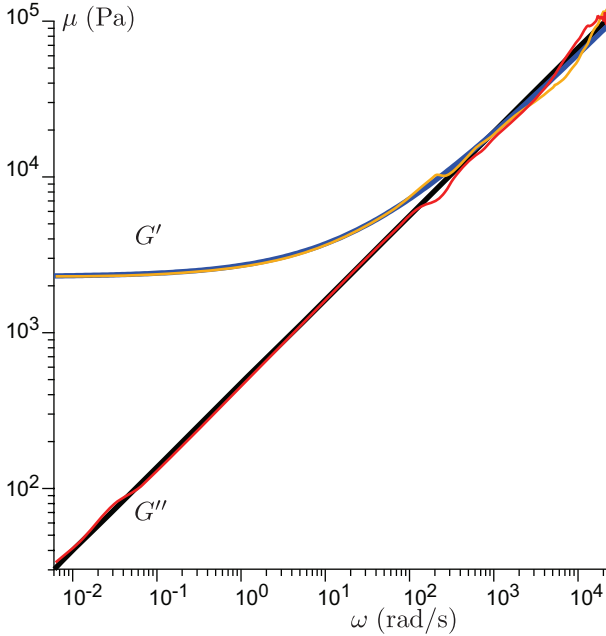}
		\caption{ Storage modulus $G'(\omega)$ and loss modulus $G''(\omega)$  as a function of the angular frequency $\omega$ for the PDMS gel. Orange and red points: measurements for a sample of PDMS gel much larger than the magnet ($R=5$ cm), deduced from the theory from $K'$ and $K''$ (Fig.~\ref{K}). Blue and black lines: fit obtained in the quasi-static domain ($\omega < 100$ rad/s) by the analytical formula $\mu=\mu_0 \left[1+(i\omega\tau)^n\right]$, here extended here to high frequencies.}
		\label{G}
\end{figure}

\textit{Determination of $\mu(\omega)$ at high frequencies~--~} In order to extract the rheology $\mu(\omega)$ from the raw data $K(\omega)$ at high frequencies, one needs to numerically solve the dynamical response in the presence of inertia. Here we present a solution strategy for incompressible media (details given in the Appendix). The key ingredient is to determine the dynamic Green's function of the system, which gives the relation between the normal stress $\sigma$ at the surface and the surface displacement $H$, for independent spatial modes. To compute the spatial modes, we introduce the stream function in cylindrical coordinates $\psi(r,z)$. Incompressibility is ensured when expressing the displacements $u_r$ and $u_z$, in terms of the streamfunction as $ u_r=-\frac{1}{r}\partial_z \psi$ and $u_z=\frac{1}{r}\frac{\partial}{\partial r} \psi$.  
Then, the modes of wavenumber $k$ are described by (see Appendix)
\begin{eqnarray}
\psi&=&[A e^{kz}+B e^{-kz}+C e^{Qz}+D e^{-Qz}]\,r J_1(kr)\nonumber\\
&{\rm with}&\quad Q\simeq k - \frac{\rho_g \omega^2}{2k\mu(\omega)}.
\end{eqnarray}
Hence the spatial modes exhibit a radial structure described by a Bessel function, while the vertical structure is a superposition of exponentials. Note the explicit dependence on the gel density $\rho_g$, expressing the inertia.
%
%
The coefficients $A$, $B$, $C$ and $D$ are determined by the boundary conditions, consisting of vanishing radial and vertical displacements on the bottom of the sample, vanishing shear stress at the free surface, outside the magnet, and imposed displacement in the contact area with the magnet (see Appendix). For an infinitely thick sample, the associated Green function $\mathcal{G}$ relating the stress to the surface displacement (for wavenumber $k$), takes the form
\begin{equation}
\mathcal{G}= \frac{\sigma}{H}=\mu \left(3 k + \frac{Q^2}{k} - \frac{4 k^2}{k + Q}\right).
\end{equation}
The final step is to find the superposition of modes that describes the disc-indenter, which requires that $H=Z$ for $r<R$ and $\sigma=0$ for $r>R$. This constitutes a standard mixed problem which can be easily solved iteratively in both directions: either one assumes that $\mu(\omega)$ is known and $K(\omega)$ is determined or the other way around.

In figure~\ref{K}, we compare the direct measurements of $K$ to the prediction of the model including inertia. We  assumed that the expression $\mu(\omega)=\mu_0\left[1+(i\omega\tau)^n\right]$ holds in whole frequency range, but the parameters are extracted from the overdamped regime (i.e. for $\omega < 100$ rad/s). The calculated value $K(\omega)$ is found to be in almost perfect agreement with the measurement. Without any adjustable parameters, one recovers the resonance frequency with a good accuracy, as well the inertial behavior at high frequency where $|K'| \sim -\omega^2$.

Figure \ref{G} shows the opposite route, where we use the inversion of the model to deduce $\mu$ from the measurement of $K$. As $K$ and $\mu$ are directly proportional in the quasi-static regime, one recovers the excellent agreement of figure~\ref{K} in the range $\omega < 100$ rad/s. However, we now have gained access to two more decades in frequency, exceeding $10^4$ rad/s. An important experimental aspect is that at large frequency, the real part of the response $K'$ is dominated by inertia: as a consequence, its dependence on $G'$ and $G''$ becomes subdominant. $K''$, on the other hand, still depends primarily on the rheology. The deduction of $\mu$ from $K$ becomes therefore more and more sensitive to the accuracy of the measurement. This requires an accurate calibration of the phase lag introduced by the vibrometer. Between $10^3$ rad/s and $10^4$ rad/s, a 5\% inaccuracy on $K$ results into a $25\%$ effect on $\mu$. Similarly, all quantities measured independently (the indenter radius $R$, its mass, the sample density $\rho$ and the calibrated constant relating $F$ to $I$) must be measured within a few per thousand accuracy to reach the percent accuracy on the rheology.

Further improvement of the method should utilize the prior assumption that the rheological response function is causal: the Fourier transform of $\mu$ must vanish at positive times. Accordingly, $\mu$ must obey the Kramers-Kronig relation, a fact that is not used here. As a consequence, the measured rheology could be improved by projecting it on the space of admissible function $\mu(\omega)$, which would compensate for the lack of information extracted from $K'$ in the inertial regime.

In conclusion, the technique is particularly interesting by its simplicity, the possibility to obtain rheological measurements in the quasi-static and in the high-frequency regime, and the possibility to perform the measurement on small samples. The absence of contact between the magnetic probe and the main part of the apparatus makes the proposed method very suitable to characterize rapidly the mechanical properties of biomedical tissues \cite{mijailovic2018characterizing}. In this context, it is particularly interesting to have a rheometer that can be brought to the patient to perform in vivo measurements. 

\newpage
\begin{widetext}
\appendix
\section{Response function}
\subsection{Axisymmetric dynamic Green's function of an incompressible medium}
We consider an incompressible layer of visco-elastic material of thickness $e$, that is characterised by a complex shear modulus $\mu(\omega)=G'(\omega)+iG''(\omega)$. The layer has vanishing displacements at the bottom. The top surface is free from stress, except on a disk of radius $R$ where an oscillatory displacement is imposed. Taking a temporal Fourier transform and considering a single mode of angular frequency $\omega$, the dynamical equation reads
 \begin{equation} 
- \rho \omega^2 \vec u=-\vec \nabla p+\mu(\omega) \vec \nabla^2 \vec u
\end{equation}
where $\vec u$ is the spatial model of the displacement field. We consider cylindrical coordinates $r,z$, and introduce the axisymmetric streamfunction $\psi$ to ensure incompressibility, defined by: 
 \begin{equation} 
 u_r=-\frac{1}{r}\partial_z \psi \quad {\rm and} \quad  u_z=\frac{1}{r}\frac{\partial}{\partial r} \psi
\end{equation}
Projecting the dynamical equation in polar coordinates, one gets:
\begin{eqnarray}
- \rho \omega^2 u_r &=& -\frac{\partial}{\partial r} p+\mu \left(\frac{1}{r}\frac{\partial}{\partial r} (r \frac{\partial u_r}{\partial r} )-\frac{u_r}{r^2}+\frac{\partial u_r}{\partial z^2} \right)\nonumber\\
- \rho \omega^2 u_z &=& -\partial_z p+\mu \left(\frac{1}{r}\frac{\partial}{\partial r} (r \frac{\partial u_z}{\partial r} )+\frac{\partial u_z}{\partial z^2} \right)
\end{eqnarray}
Eliminating pressure between the two equations, we get an equation on the stream function:
 \begin{eqnarray} 
\Delta \left(\rho \omega^2 +\mu  \Delta\right) \psi=0 \nonumber\\
{\rm with}\quad \Delta=\left(r \frac{\partial}{\partial r} \left(\frac{1}{r} \frac{\partial}{\partial r}\right)  + \frac{\partial}{\partial z^2}\right) 
\end{eqnarray}
Spatial modes can be related to Bessel functions, noting that  the differential equation $r (f'(r)/r)'=-k^2 f(r)$ has as solutions, $r J_1(kr)$ and $r Y_1(kr)$. Along the direction normal to the surface, the equations are homogeneous. The vertical structure of a mode is therefore a superposition of exponentials of decay rate $q$, which satisfy the equation
 \begin{equation} 
(q^2-k^2) \left(\rho \omega^2 +\mu(\omega) (q^2-k^2)\right) \psi=0.
\end{equation}
Solutions are $q=\pm k$ or $q=\pm Q$ with
 \begin{equation} 
Q^2=k^2 - \kappa^2\quad {\rm with} \quad \kappa^2= \frac{\rho \omega^2}{\mu(\omega)},
\end{equation}
where by continuity, at small $\kappa$, the root must obey:
 \begin{equation} 
Q\simeq k - \frac{\kappa^2}{2k}.
\end{equation}
%
The solution for $\psi$ therefore reads:
 \begin{eqnarray} 
\psi&=&[A_j\exp(kz)+B_j\exp(-kz)+C_j \exp(Qz)+D_j \exp(-Qz)]\,r J_1(kr)\\
&+&[A_y\exp(kz)+B_y\exp(-kz)+C_y \exp(Qz)+D_y \exp(-Qz)]\,r Y_1(kr)
\end{eqnarray}
where the constants are set by the boundary conditions. The pressure field is obtained by integration of $\frac{\partial}{\partial r} p$.
\begin{equation}
p=p_0+\kappa^2 \left(A_j\exp(kz)-B_j\exp(-kz))J_0(kr)+ (A_y\exp(kz) -B_y\exp(-kz))Y_0(kr)\right)
\end{equation}
At the bottom of the layer, we impose a vanishing displacement $u_r=0$ and $u_z=0$:
\begin{eqnarray}
u_r(z=-e)&=& (A_j k e^{-ke}-B_j k e^{ke}+C_j Q e^{-Qe}-D_j Q e^{Qe}) J_1(kr)\\
&+&(A_y k e^{-ke}-B_y k e^{ke}+C_y Q e^{-Qe}-D_y Q e^{Qe}) Y_1(kr)=0\\
u_z(z=-e)&=&[A_je^{ke}+B_je^{-ke}+C_j e^{Qe}+D_j e^{-Qe}]\,r J_0(kr)\\
&+&[A_ye^{ke}+B_ye^{-ke}+C_y e^{Qe}+D_y e^{-Qe}]\,r Y_0(kr)=0.
\end{eqnarray}
At the free surface, located at $y=0$, we impose a null vanishing stress $\sigma_{xy}=0$ and we want to determine the disturbance to the normal stress $\sigma_{yy}$. The condition for the tangential stress reads:
\begin{equation}
\frac{\partial}{\partial r} u_z+ \partial_z u_r=-(2(A_j +B_j)k^2+(C_j+D_j) (k^2+Q^2))J_1(kr)-(2(A_y +B_y)k^2+(C_y+D_y) (k^2+Q^2))Y_1(kr)=0.
\end{equation}
Using regularity of the solution in $r=0$, the second Bessel function which diverges at the origin, must be excluded. The three boundary conditions then reduce to:
\begin{eqnarray}
2(A_j +B_j)k^2+(C_j+D_j) (k^2+Q^2)&=&0\nonumber\\
(A_j k e^{-ke}-B_j k e^{ke}+C_j Q e^{-Qe}-D_j Q e^{Qe}) J_1(kr)&=&0\nonumber\\
(A_je^{ke}+B_je^{-ke}+C_j e^{Qe}+D_j e^{-Qe})\,r J_0(kr)&=&0\nonumber.
\end{eqnarray}
Now, we wish to relate the normal stress at the free surface, $\sigma = \sigma_{yy}(y=0)$, to the surface displacement $h= u_y(y=0)$:
\begin{eqnarray}
\sigma&=&\sigma_{zz}(z=0)=2 \mu \partial_z u_z - p\\
&=&\mu \left(2 k ((A_j-B_j)k +(C_j-D_j) Q) -\kappa^2 (A_j-B_j)\right)J_0(kr),\nonumber
\end{eqnarray}
and
\begin{equation}
Z=u_z(z=0)=(A_j+B_j+C_j+D_j) k J_0(kr).
\end{equation}
Eliminating the coefficients $A_j$, $B_j$, $C_j$ and $D_j$, we obtain the green function:
\begin{eqnarray}
\frac{\sigma}{\mu Z}=  \frac{Q \left(5 k^4+2 k^2 Q^2+Q^4\right) \cosh (e k) \cosh (e Q)}{k (k^2-Q^2) (k \cosh (e k) \sinh (e Q)-Q
   \sinh (e k) \cosh (e Q))}\\
 - \frac{k \left(\left(k^4+6 k^2 Q^2+Q^4\right) \sinh
   (e k) \sinh (e Q)+4 k Q \left(k^2+Q^2\right)\right)}{k (k^2-Q^2) (k \cosh (e k) \sinh (e Q)-Q
   \sinh (e k) \cosh (e Q))}
\end{eqnarray}
In the limit where $e$ goes to infinity, the Green function reduce to:
\begin{eqnarray}
\frac{\sigma}{Z}&=& \mu \left(3 k + \frac{Q^2}{k} - \frac{4 k^2}{k + Q}\right) \\
&=&\mu \left(4 k - \frac{\kappa^2}{k} - \frac{4 k^2}{k + Q}\right).
\end{eqnarray}
This equation is presented in the main text, and forms the basis for the numerical inversion from the force measurement to the rheology $\mu(\omega)$.

\subsection{Asymptotic expansions}
When inertia is negligible, $Q$ tends to $k$ and the Green's function takes the limiting form:
\begin{equation}
\frac{\sigma}{Z}=  \mu \frac{2 k \left(2 e^2 k^2+\cosh (2 e k)+1\right)}{\sinh (2 e k)-2 e k} 
\end{equation}
At small $e/R$, we can simplify further by taking the limit $ek\to0$ which coincides with the lubrication approximation:
\begin{equation}
\frac{\sigma}{Z}=  \frac{3\mu}{k^2 e^3} 
\end{equation}
In real space, this gives back the equation presented in the main text
\begin{equation}
\frac{\partial^2 \sigma}{\partial r^2}+\frac{1}{r}\;\frac{\partial \sigma}{\partial r}=\frac{3\mu}{e^3}Z(r)
\end{equation}
which gives for $\sigma$.
\begin{equation}
\sigma=\frac{3\mu}{4e^3}(r^2-R^2) h.
\end{equation}
The force therefore gives the result presented in the main text:
\begin{equation}
\frac{F}{ZR}=\frac{3\pi \mu R^3}{8e^3}.
\end{equation}
%
%

At large $\omega$, conversely, we expect
\begin{equation}
\sigma=- \frac{\rho \omega^2}{k}  Z
\end{equation}
which will therefore lead to the scaling law:
\begin{equation}
\frac{F}{ZR}\sim \rho \omega^2 R^2,
\end{equation}
as is evidenced also in our experiments.

\subsection{Discrete Hankel transform}
Numerically, we use the discrete Hankel transform, defined by:
\begin{equation}
f(r)=\sum_{n=0}^{N-1} \hat f_n J_0(\alpha_n r)
\end{equation}
where the discrete eigenvectors $\alpha_n$ denotes the $n$th root of Bessel function ($\alpha_n\simeq 3\pi/4+n\pi$ at large $n$). To project the continuous equations, we evaluate them at $N$ discrete values of $r$, labelled $r_k$ and defined by:
\begin{equation}
r_k=\frac{\alpha_k}{\alpha_N}
\end{equation}
so that:
\begin{equation}
f(r_k)= f_k=\sum_{n=0}^{N-1}  \frac{2}{\alpha_N J_1^2(\alpha_n)}  \hat f_n J_0\left(\frac{\alpha_k \alpha_n}{\alpha_{N}}\right)
\end{equation}
and
\begin{equation}
\hat f_n=\sum_{k=0}^{N-1}  \frac{2}{\alpha_N J_1^2(\alpha_k)} f_k J_0\left(\frac{\alpha_k \alpha_n}{\alpha_{N}}\right)
\end{equation}
The mixed problem, defined by an imposed displacement $Z(r)$ normalised to $1$ for $r<R$ and a vanishing normal stress $\sigma(r)$ for $r>R$, is solved using this representation for both the displacement $Z(r)$ and the stress $\sigma(r)$. The Hankel transform is used to compute $\sigma$, when $Z$ is known, and reciprocally, using the Green function in the reciprocal space. The algorithm is iterative. At each stage, $Z$ is imposed to be unity for $r<R$ but is kept as it is for $r>R$. The associated stress is determined, which is set to $0$ for $r>R$. The new test profile $Z(r)$ is then determined using the Green function backward. When this simple algorithm does not converge, a small factor $\epsilon$ is introduced to superimpose the old profile, weighted by $1-\epsilon$ and the new one, weighted by $\epsilon$.

The force on the indenter is measured as: 
\begin{equation}
F=2\pi \int_0^{2\pi} r \sigma_{yy} (r,0) dr
\end{equation}
using the integral relation:
\begin{equation}
\int_0^R 2\pi r f(r) dr=\sum_{n=0}^{N-1}  \frac{2R}{\alpha_n \alpha_N J_1^2(\alpha_n)}  \hat f_n J_1\left(\alpha_n R\right)
\end{equation}
\end{widetext}

%

\end{document}